\documentclass[sigconf,bookmarksnumbered,unicode,bookmarks=false,balance=true,hidelinks,review=false,timestamp=false]{acmart}

\usepackage{tikz}
\newcommand*\circled[1]{\tikz[baseline=(char.base)]{
            \node[shape=circle,draw,inner sep=2pt] (char) {#1};}}

\usepackage{titlesec}
\titlespacing*{\subsection}{0pt}{0.6\baselineskip}{0.2\baselineskip}
\titlespacing*{\subsubsection}{0pt}{0.6\baselineskip}{0.2\baselineskip}

\pdfoutput=1
\usepackage[utf8]{inputenc}
\usepackage[]{numprint}
\usepackage[inline]{enumitem}
\newcommand*\np[2][z]{
\ifx z#1%
$\numprint{#2}$%
\else%
$\numprint[#1]{#2}$%
\fi\xspace%
}

\copyrightyear{2024}
\acmYear{2024}
\setcopyright{rightsretained}
\acmConference[ICSE-NIER'24]{New Ideas and Emerging Results }{April 14--20, 2024}{Lisbon, Portugal}
\acmBooktitle{New Ideas and Emerging Results (ICSE-NIER'24), April 14--20, 2024, Lisbon, Portugal}\acmDOI{10.1145/3639476.3639764} \acmISBN{979-8-4007-0500-7/24/04}

\usepackage{pifont}

\begin{document}

\title{Breaking the Silence: the Threats of Using LLMs in Software Engineering}

\author{June Sallou}
\affiliation{%
    \institution{TU Delft}
    \country{The Netherlands}
}
\email{J.Sallou@tudelft.nl}
\orcid{0000-0003-2230-9351}

\author{Thomas Durieux}
\affiliation{%
    \institution{TU Delft}
    \country{The Netherlands}
}
\email{thomas@durieux.me}
\orcid{0000-0002-1996-6134}

\author{Annibale Panichella}
\affiliation{%
    \institution{TU Delft}
    \country{The Netherlands}
}
\email{A.Panichella@tudelft.nl}
\orcid{0000-0002-7395-3588}

\renewcommand{\shortauthors}{Sallou, et al.}

\begin{abstract}
Large Language Models (LLMs) have gained considerable traction within the Software Engineering (SE) community, impacting various SE tasks from code completion to test generation, from program repair to code summarization. Despite their promise, researchers must still be careful as numerous intricate factors can influence the outcomes of experiments involving LLMs. 
This paper initiates an open discussion on potential threats to the validity of LLM-based research including issues such as closed-source models, possible data leakage between LLM training data and research evaluation, and the reproducibility of LLM-based findings.
In response, this paper proposes a set of guidelines tailored for SE researchers and Language Model (LM) providers to mitigate these concerns.
The implications of the guidelines are illustrated using existing good practices followed by LLM providers and a practical example for SE researchers in the context of test case generation.
\end{abstract}


\maketitle

\vspace{-0.5em}
\section{Introduction}
In recent years, the utilization of Large Language Models (LLMs) has gained substantial traction within the Software Engineering (SE) community. Equipped with language understanding and generation capabilities, these models have impacted various SE research and practice aspects. From code generation to bug detection and natural language interactions with codebases, LLMs have played a pivotal role in recent SE advancements~\cite{Ozkaya2023Apr, 10.1109/ICSE48619.2023.00085, Fan2023May, xia2023automated, siddiq2023exploring}.

Despite their promise, researchers must tread cautiously when making claims about the effectiveness of their approaches. The outcomes of experiments involving LLMs can be influenced by numerous intricate factors that can be challenging to discern or control. This intricacy underscores the need to thoroughly examine the validity of research findings when LLMs are involved.

\begin{figure}
    \centering
    \includegraphics[width=0.38\textwidth]{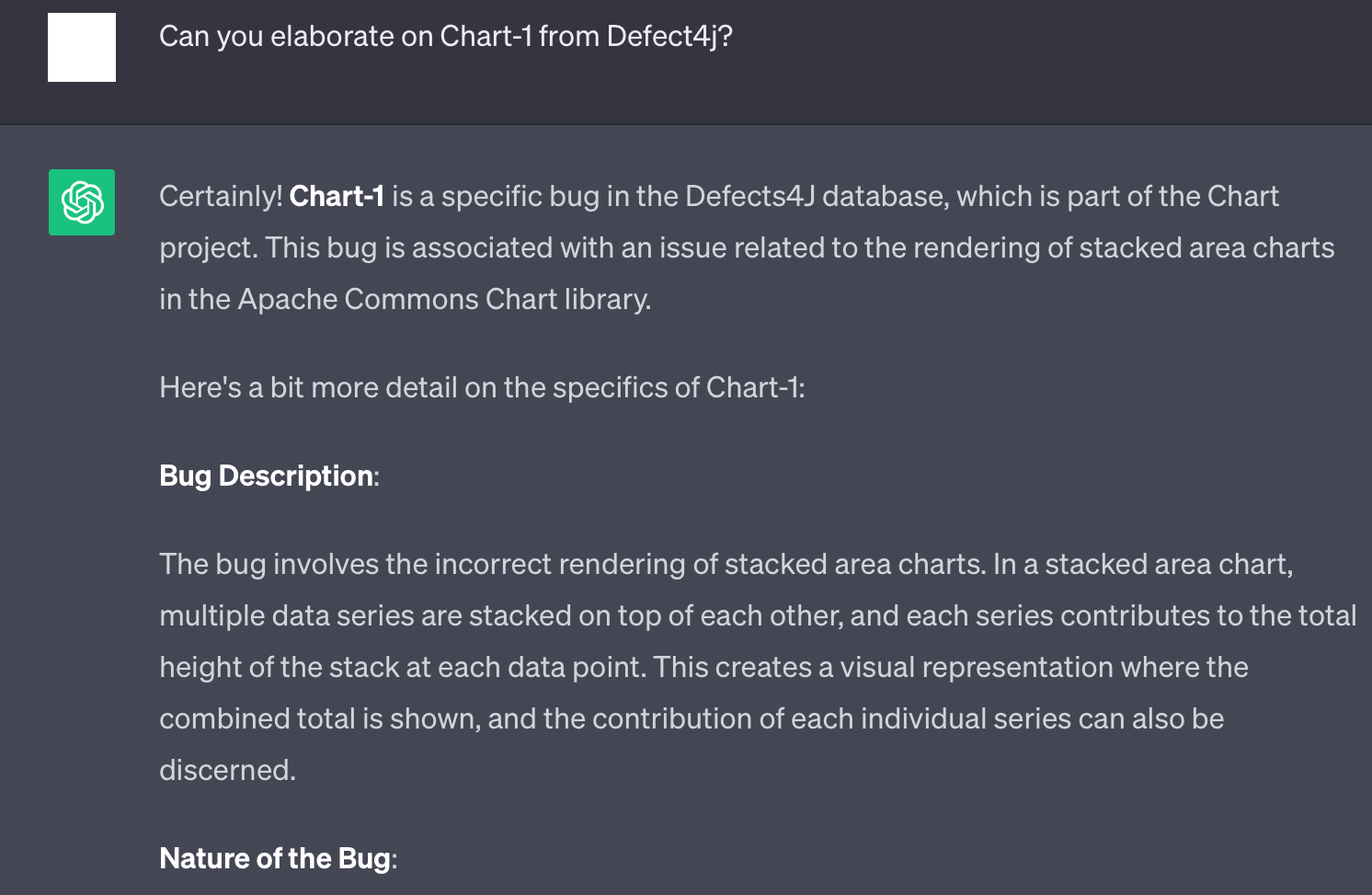}
    \caption{ChatGPT's detailed answers about a specific bug in Defects4J.}
    \label{fig:defects4j_chatgpt}
\end{figure}

For instance, consider the evaluation of LLMs using well-known projects such as those included in Defects4J~\cite{just2014defects4j}. Although ChatGPT-3.5, as a general-purpose chatbot, is not fine-tuned for specific SE tasks, it possesses precise knowledge of the bugs within these projects (see Figure~\ref{fig:defects4j_chatgpt}). Thus, when OpenAI models are employed for tasks like patch generation~\cite{xia2023keep}, fault localization~\cite{wu2023large}, or test generation~\cite{tang2023chatgpt} for Defects4J, they have superior performance compared to an unknown code~\cite{siddiq2023exploring}.
This foreknowledge arises from the pre-trained process, as ChatGPT has been trained on a large variety of datasets, including scientific papers (though the specifics are not fully disclosed). This raises severe concerns about the threats to both \textit{construct} (\textit{training and evaluating on the same dataset}) and \textit{external} (\textit{do the results hold for unknown projects/code?}) validity.

This paper aims to initiate a community-wide discussion and raise awareness of these issues to facilitate collective progress. Specifically, we focus on three key threats to validity:
\circled{1} Using \textit{closed-source} LLMs and their implications w.r.t. data \textit{privacy} and the models' \textit{evolution} unpredictability.
\circled{2} The \textit{blurry separation} between training, validation, and test sets and the corresponding potential explicit/implicit data leakage.
\circled{3} \textit{Reproducibility} of the published research outcomes over time, due to the \textit{non-stochastic} nature of LLMs answers, the non-transparent releases of \textit{new model versions}, and the lack of complete \textit{traceability}.

While recent papers acknowledge some of these concerns (e.g.,~\cite{ye2023assessing, jesse2023large, siddiq2023exploring}), we highlight the necessity for further empirical methodologies—such as code obfuscation, multiple independent prompts or queries, and metadata provision—to alleviate and address these concerns. Therefore, we present an initial set of guidelines aimed at mitigating these threats, specifically targeting SE researchers and Language Model (LM) providers.
While we provide a list of actionable suggestions, we emphasize the importance of the wide SE community to follow our initial steps and build a more comprehensive list of threats to validity and research methods to address them, ultimately advancing the field while ensuring the reliability and validity of LLM-based research contributions.

To illustrate the practical application of our guidelines, we demonstrate their implementation in the context of test case generation ---a well-established SE task~\cite{fraser2011evosuite,panichella2015reformulating}--- using ChatGPT 3.5 on two buggy code snippets from Defects4J~\cite{just2014defects4j}. We provide a full replication package~\cite{repo-expe} including (1) the implementation of our guidelines, and (2) the results of our analysis. Finally, we highlight existing good practices from LLMs providers that align with our guidelines.


\section{Threats to Validity for LLMs}

This section opens the discussion related to LLM-based research within the SE research community. 

\subsection{Closed-Source Models} \label{closedsourcethreat}
A significant portion of the LLMs community heavily relies on closed-source models, as evidenced by the prevalence of ChatGPT-related papers at this year's ICSE conference. 
This dependence stems from factors such as the models' effectiveness, availability, and cost-effectiveness. To put this into perspective, deploying an open-source LLM model comparable to ChatGPT, like the Falcon 180B model, would incur a monthly cost of \$29,495 on AWS.\footnote{Visited on 7-Dec-2023, 8x A100 80GB on \href{aws}{https://aws.amazon.com/ec2/instance-types/}}

However, the utilization of closed-source models introduces significant threats to the validity of the research approach.

\textbf{Model Evolution Unpredictability.} One primary concern arises from the lack of control over the evolution of these closed-source models. New models may be released into production~\cite{pozzobon2023challenges}, and notable changes in the output of OpenAI models have been observed~\cite{Chen2023Jul}. Such changes can occur during or after the research approach has been presented, potentially making the presented results obsolete. 
Moreover, this concern is also particularly pronounced for incremental works that use LLM. Indeed, distinguishing whether the improvements claimed in the new contribution are the result of changes to the LLMs' models or due to the novelty of the contribution becomes a complex task.
    
\textbf{Privacy Implications.} Another significant aspect of concern is privacy. Closed-source models often lack transparency, making it difficult to assess the privacy implications associated with their usage~\cite{Al-Kaswan2023Feb} as well as potential copyright infringements. 

\subsection{Implicit Data Leakage}


LLMs are trained on vast textual datasets such as Wikipedia, GitHub, and StackOverflow~\cite{touvron2023llama, chen2021evaluating} and derive their understanding of semantics and contextual word relationships from this diverse data. These models contain millions (e.g., BERT) to billions (e.g., ChatGPT-4 and LLaMA) of parameters, which undergo a series of iterative optimizations during \textit{pre-training} that minimize the loss function.

\textbf{Data leakage due to pre-training}. 
Pre-training in \textit{unsupervised} or \textit{semi-supervised} learning does not tailor models for specific software engineering tasks they will be later evaluated on after parameter re-tuning. 
Nonetheless, questions can be raised whether LLMs potentially memorize existing code samples used for evaluation, instead of generating new, unseen code~\cite{inan2021training}.
For instance, prior studies~\cite{ asare2022github, pearce2023examining, jesse2023large, Karmakar2022Dec} highlighted vulnerabilities in code generated by Codex~\cite{chen2021evaluating}, originating from its training set.
Siddiq et al.~\cite{siddiq2023exploring} investigated the performances of three LLMs (Codex, ChatGPT3.5~\cite{openai}, and StarCoder~\cite{nijkamp2022codegen}) in generating unit tests for Java programs. They reported remarkable performance discrepancies between the HumanEval~\cite{athiwaratkun2022multi} ($>$69\% branch coverage) and the SF110~\cite{fraser2011evosuite} (2\% branch coverage) datasets. We remark that the former is available on GitHub~\cite{athiwaratkun2022multi} while the latter is not (available on \texttt{SourceForge} instead).

Additionally, previous studies in \textit{metamorphic testing}~\cite{compton2020embedding, yefet2020adversarial, applis2023searching, yang2022natural} have demonstrated how semantic-preserving changes to code snippets can effectively deceive LLMs. These approaches create new data points that differ from the original ones by at least one metamorphic change, increasing the likelihood that LLMs will not recognize the code snippets seen during pre-training.

\textbf{Data leakage due to fine-tuning}.
LMs applied to specific SE tasks require parameter tuning via \textit{supervised} learning, adjusting parameters using labeled datasets specific to the task. Despite efforts to separate training, validation, and test sets, ensuring clear distinctions is not guaranteed. 
In practice, different projects might have common dependencies and use the same pre-defined APIs. 
For instance, within the Java ecosystem, numerous libraries/tools often rely on common dependencies like Log4j, Apache Commons, Spring, GSon, and others. Consequently, a scenario of data leakage can arise if the LLM is trained on ``project A'' that employs a specific API, and the resulting model is subsequently used to fix the usage of the same API in another project within the test set.


\subsection{Reproducibility}

Several concerns can be raised w.r.t the reproducibility of the LLMs outputs. The ability to obtain identical results following the same procedure by external parties is proven to be challenging.

\textbf{Output Variability.} LLMs exhibit variability in their outputs, even when using identical input.
Running the same prompt several times may not result in identical output, rendering the usage of LLMs non-deterministic.
Examples of such phenomena have been described in the literature, including other application domains, such as the medical domain~\cite{Epstein2023Jul}. 
We also experimented with an example in software engineering involving a code generation task. 
In this use case, we demonstrate that running a prompt to generate Python code multiple times results in different responses from the GPT-3.5 model. To avoid data contamination, we get inspiration from the methodology outlined by Chen et al.~\cite{Chen2023Jul}. We use a coding challenge from the LeetCode~\cite{LeetCode} platform as the prompt, employing the same prompt twice in two separate sessions\footnote{first session: \url{https://chat.openai.com/share/a0c7ef5c-74ce-466b-a1d5-5f44e03a626d}, second session:\url{https://chat.openai.com/share/6566acff-12eb-470a-a043-3e2294cf6406}} on the same day (September 12th, 2023). During these sessions (whose chat links are provided as footnotes), we observe distinct codes generated by GPT-3.5, with varying function operations reasoning, variable names, and initialization values or expressions.

\textbf{Time-Based Output Drift.} Furthermore, there is no assurance that the results will remain consistent over time. 
As discussed in Section~\ref{closedsourcethreat}, many LLMs are closed-source, and there are no established practices akin to regression testing to account for output variability.  
Running the same prompt at a later time (e.g., days or months) may lead to a drift in the outcome due to potential retraining between sessions, reinforcement learning between sessions, or adjustments based on user feedback.
Chen et al.~\cite{Chen2023Jul} explored this time-based output drift in terms of accuracy for two versions of GPT models over a three-month interval. They show that, over a range of tasks and application domains, the overall accuracy of outputs decreases, accompanied by a remarkable mismatch in answers. In particular, for code generation, the mismatch is evaluated at 20\% for GPT-3.5 and 50\% for GPT-4 between March 2023 and June 2023. Moreover, the number of executable outputs drops from 52\% to 10\% for GPT-3.5, and from 22\% to 2\% for GPT-4.0 for the same period.

\textbf{Traceability.} Another critical concern associated with the widespread adoption of LLMs is the lack of traceability. Currently, connecting the output of LLMs to specific prompts, along with 'configuration' details such as the version of the used LLM, the date of the query, and other specifications, can be a challenging task.



\section{Discussion and Guidelines}
In this section, we present initial guidelines and methodologies addressing the mentioned threats. While we provide a list of actionable suggestions as opening steps, 
we encourage the SE community to work toward establishing standards and expectations at the same level as those commonly applied with traditional AI techniques. We organize the guidelines according to the actors they involve: the LLM providers, and the SE researchers using LLMs.

\subsection{Guidelines for LLM Providers}

\subsubsection{Closed-Source Models} We foresee two main strategies to address this category of threats to validity:

\textbf{Enhance model transparency.} LLM providers should prioritize transparency by furnishing comprehensive information about their models. This should encompass details on the model's creation process and the methodology used for data selection during training. Furthermore, providers should share statistics and data-point information to shed light on the model's training dataset. Ideally, an API service could be established, enabling users to verify if a particular data source was included in the model's training or validation datasets. Such a service would not only enhance transparency but could also serve privacy and copyright verification purposes.

\textbf{Use versioning information.} Providers should provide their model version, and they should adopt a versioning nomenclature that distinguishes major revisions from minor updates. This enables users to discern the significance of model version changes.

\subsubsection{Data Leakage}
In light of concerns regarding closed-source models, LLM providers should provide services that allow researchers to verify which projects and sources were considered during pre-training. A positive example is CodeBERT, whose provides do not disclose pre-training code but enable verification of included projects for pre-training through the train split data.\footnote{\url{https://huggingface.co/datasets/code_search_net}}

\subsubsection{Reproducibility} We propose two methods to address this:
\noindent \textbf{Use a fixed random seed.} 
LLM providers should ensure the inclusion of a settable random seed during the inference of LLMs,  render the inference deterministic for each specific case. 
This practice would help address the variability of output concerning traceability and reproducibility. In the case of closed-source LLMs, a dedicated API could be made accessible, allowing the user to set the seed without requiring access to the entire model. Toward this direction,
OpenAPI has recently released a \textit{beta feature}~\footnote{\url{https://platform.openai.com/docs/guides/text-generation/reproducible-outputs}} that allows users to set fixed seeds during prompting, although deterministic answer is not fully guaranteed due to different back-end settings.

\textbf{Use an archiving system.} In addition to the Versioning Information, we advocate for the usage of a general archiving system, to ensure that external parties can reproduce the observations made by LLMs. 
It should be noted that some efforts are already being made in that direction. We can cite the HuggingFace platform~\cite{HuggingFace}, which provides pre-trained models for download with information about the model training, file versioning, and datasets. Zenodo~\cite{Zenodo} is another example of storing and making versioned models and datasets accessible and reusable, which is commonly used in the research community. 
However, the use of such platforms is not yet a regular and consistent practice. Moreover, a dedicated LLM platform is still missing, as LLM sizes are generally large, posing challenges for downloading or uploading

\subsection{Guidelines for SE Researchers}
This section outlines guidelines for SE researchers. Along with presenting guidelines, we exhibit their practical applicability through a showcase example, using ChatGPT3.5 to generate JUnit test cases for two buggy programs in the Defect4j dataset~\cite{just2014defects4j}, \texttt{Chart-11} and \texttt{Math-5}. The prompts with the collected answers, data analysis, and metadata are available in our replication package~\cite{repo-expe}.

\subsubsection{Reproducibility}

To address the threats to the reproducibility of LLMs-based approaches, we proposed the following guidelines: 

\textbf{Assess output variability.} Due to output variability, running the LLMs' inference only once is insufficient to ensure reproducibility. Therefore, we argue in favor of conducting multiple replication runs and using variability metrics during the evaluation. For our showcase example, we queried ChatGPT3.5 ten times over different days using the same prompts (see our replication package) and targeting only the buggy methods (i.e., no the entire classes). We then analyzed the resulting branch coverage and test execution results. For \texttt{Math-5}, we report an average branch coverage of 70\% for the tested Java method with a large variability (interquartile range or IQR) of 27.5\%. We also observe variability in the number of generated tests (between 5 and 7) and number of failing tests (between 1 and 2). 
For \texttt{Chart-11}, the generated tests achieve 71\% of branch coverage for the tested Java method, with 20\% IQR. ChatGPT also generates between 1 and 5 failing tests for this method.

\textbf{Provide execution metadata.} Associated with the LLMs inference results, we argue that relevant additional data should be made accessible and considered during the evaluation of LLMs. Such information includes, but is not limited to:
(i) \textit{Model information:} To reproduce the LLMs' results and evaluation, information concerning the model is necessary (e.g., version, seed, etc.). Furthermore, the model itself should be accessible to enable its use, at least in a black box manner.
(ii) \textit{Prompts:} The exact inputs (queries) used for the inference and evaluation of the LLMs.
(iii) \textit{Date of LLMs query:} The date is relevant data to share to address the time-based output drift (that can happen because of retraining of models, or reinforcement learning from past human feedback and interactions).
(iv) \textit{Output variability metrics and associated assessment package:} Information concerning the assessment of output variability would enable the user to understand the risk regarding consistency in using the LLM in question. Providing the package containing the prompts and results used during this evaluation would help to ensure reproducibility.
(v) \textit{Scope of reproducibility:} To ensure the trusted and controlled usage of LLMs, information about the scope in which the model has been trained or assessed should be disclosed, including the application domains and studied use cases.

We provide an example of metadata (written using the JSON format) for the showcase of this paper in our replication material.

\subsubsection{Data Leakage}
A few recommendations can be made to tackle the crucial concerns about the potential data leakage:

\textbf{Assess LLMs on metamorphic data}.
\textit{Metamorphic testing} is active research for the model robustness~\cite{compton2020embedding, yefet2020adversarial, applis2023searching}. Metamorphic testing generates new data samples (code) by applying metamorphic transformations to the validation or test sets. These new snippets maintain semantic and behavioral equivalence with the original code, yet exhibit structural differences (e.g., distinct Abstract Syntax Trees).
Prior studies have shown that \texttt{CodeBERT} and \texttt{code2vec} are not robust, i.e., they produce different (worse) results when obfuscating the variable names~\cite{compton2020embedding}, introducing unused variables~\cite{yefet2020adversarial}, replacing tokens with plausible alternatives~\cite{cito2022counterfactual}, and wrapping-up expressions in identity-lambda functions~\cite{applis2023searching, applis2021assessing}.

Therefore, we advise researchers to complement the analysis of the LLMs performance with new data samples generated with metamorphic testing. 
The selection of metamorphic transformations should align with the specific task at hand.  
For example, code obfuscation (for method/class names) should not hinder the ability of LLMs to generate unit tests or patches successfully. Instead, identifier names are crucial for NLP-related tasks (e.g., method name prediction), and other metamorphic transformations should be applied (e.g., wrapping up expressions in identity-lambda functions). 

To show the practicability of this guideline, we have applied code metamorphic transformations to the Java methods of \texttt{Math-5} and \texttt{Chart-11}. In particular, we (1) removed the \texttt{javadoc} and  (2) renamed the method under test and its input parameters. For renaming, we did not use randomly generated string but opted for synonyms and English words (e.g., changing the method name \texttt{reciprocal()} in~\texttt{complementary()} for \texttt{Math-5}) to maintain the naturalness of the transformed code~\cite{yang2022natural}. While we do not observe a significant difference in terms of branch coverage achieved for \texttt{Chart-11} ($p$-value=0.79 according to the Wilcoxon test) between the original and transformed code, we report a \texttt{small} negative effect size ($\hat{A}_{12}$=0.63) w.r.t. the number of failing tests (larger for the transformed code). The difference we obtained for \texttt{Math-5} on the obfuscated code is much more significant. While ChatGPT constantly generated tests for the original program with a branch coverage of 71\%, it struggled to generate any meaningful tests on the transformed code. In particular, ChatGPT always created non-compiling tests with clear examples of \textit{hallucination}~\cite{siddiq2023exploring,zhang2023siren}, i.e., invoking methods/constructors that were never included in the prompts.

\textbf{Use different sources}. As shown by Siddiq et al.~\cite{siddiq2023exploring}, LLMs achieve much worse results on projects from SourceForge compared to GitHub. Hence, we recommend researchers gather software projects and data from multiple sources.

\textbf{Code clone detection}. Given the current low transparency of closed-source LLMs, tracing the projects used for pre-training is challenging, if not impossible. However, researchers can use well-established code clone detection techniques~\cite{ain2019systematic} to check if the generated code (e.g., test cases) is similar to code seen in online repositories (e.g., manually-written test cases).

\textbf{Check for common dependencies}.
To prevent implicit data leakage between training, evaluation, and test sets (for task-specific evaluation), researchers should (1) cross-compare the external dependencies between projects belonging to different sets, (2) use code cloning techniques to assess whether similar code (e.g., API uses) appear between different projects from the different sets.
    
\subsubsection{Closed-Source Models}

\textbf{Perform comparative analysis.} Researchers are encouraged to run experiments using both open-source and closed-source LLMs. 
This comparative approach can provide additional insights into the strengths and limitations of each. Notably, the open-source LLM community has witnessed an expansion in availability, with models like llama2 \cite{touvron2023llama2} and Falcon 180B \cite{falcon} emerging as viable options. llama2 models, in particular, offer the advantage of running on consumer-grade devices.

\textbf{Framework Facilitation.} To streamline the evaluation process across multiple models, researchers can leverage frameworks such as ONNX \footnote{\url{https://onnx.ai/}}. ONNX simplifies the transition between various models, enhancing the efficiency and consistency of experimentation.

    





\vspace{-0.1em}

\section{Conclusion and Future Work}

In this article, we have initiated a discussion about the usage of LLMs in SE contributions, along with the challenges and threats to validity they bring. We have identified three primary challenges: the reliance on closed-source LLMs, potential data leakages, and concerns about reproducibility. To mitigate these, we propose an initial set of guidelines. We aim to encourage an ongoing dialogue within the community to navigate these challenges effectively.

It is essential to continue reflecting and staying critical about the usage of LLMs in SE. We must collectively define guidelines and expectations within our community. 
We believe the conversation should be spread by organizing panels with different experts and stakeholders. We should also monitor the evolution of good practices in the literature and maintain community guidelines. 

We emphasize the importance of disseminating evaluation expectations from the SE to the ML community, fostering mutual understanding of evolving practices. It's noteworthy that metrics drawn from Natural Language Processing (NLP) studies (e.g., counting the number of compiling unit tests generated by LLMs as done in~\cite{tufano2020unit, alagarsamy2023a3test}) do not adequately reflect the well-established performance metrics for SE tasks that do not have an NLP focus (e.g., see the existing standards for assessing unit test generation tools~\cite{panichella2021sbst, arcuri2014hitchhiker, JUGE}).

We strongly believe that breaking the silence as a community will enhance the validity and reliability of our LLM-based contributions and the SE community in general. The goal is to ultimately advance the field while ensuring high research quality and high methodological standards (considering different aspects, including data privacy~\cite{Al-Kaswan2023Feb}, carbon footprint~\cite{SLRGreenAI}, etc.).


\clearpage

\balance
\bibliographystyle{ACM-Reference-Format}
\bibliography{references}
\balance

\end{document}